\documentclass[fleqn,10pt]{wlscirep}
\usepackage{graphicx}
\usepackage{amsfonts}
\usepackage{color}
\usepackage{amsmath}
\usepackage{hyperref}

\newcommand{\textgx}[1]{\textcolor{black}{#1}}

\title{Differential Variance Analysis of soft glassy materials: a direct method to quantify and visualize dynamic heterogeneities}

\author[1,2,3,*]{Raffaele Pastore}
\author[4]{Giuseppe Pesce}
\author[3]{Marco Caggioni}

\affil[1]{CNR--SPIN, sezione di Napoli, Via Cintia, 80126 Napoli, Italy}
\affil[2]{University of Cincinnati, UC Simulation Center, 2728 Vine Street, Cincinnati,  OH 45219, USA}
\affil[3]{Corporate Engineering, The Procter \& Gamble Company, Cincinnati, 8256 Union Centre Blvd., West Chester, OH 45069, USA}
\affil[4] {Universit\`a di Napoli ``Federico II'', Dipartimento di Fisica, Via Cintia, 80126 Napoli, Italy}
\affil[*]{r.pastore.phys@gmail.com}

\begin{abstract}
Many amorphous materials show spatially heterogenous dynamics, as
different regions of the same system relax at different rates. 
Such a signature, known as Dynamic Heterogeneity, has been crucial to understand the jamming transition 
in simple model systems and, currently, is considered very  promising to characterize more complex fluids of industrial and biological relevance.
Unfortunately, measurements of dynamic heterogeneities
typically require sophysticated experimental set-ups and are performed by few specialized groups. 
It is now possible to quantitatively characterize the relaxation process and the emergence of dynamic heterogeneities
using a straightforward method, here validated on video microscopy data of hard-sphere colloidal glasses.
We call this method Differential Variance Analysis (DVA), since
it focuses on the variance of the differential frames, obtained subtracting images at different lag-times.
Moreover, direct visualization of dynamic heterogeneities naturally appears in
the differential frames, when the lag-time is set to the one corresponding to the maximum dynamic susceptibility.
This approach opens the way to effectively characterize and tailor a wide variety of soft materials, 
from complex formulated products to biological tissues.
 
 \end{abstract}

\begin{document}

\flushbottom
\maketitle

Many complex fluids, when changing control parameters like temperature or composition, exhibit a jamming transition from a 
liquid-like to an amorphous solid-like state. Approaching such transitions, the dynamics dramatically slows down
and shows increasing spatio-temporal fluctuations, known as Dynamic Heterogeneities (DHs)~\cite{DHbook}.
This dynamic signature is especially relevant for glass forming systems, 
such as supercooled liquids and dense colloidal suspensions:
since the glass transition has been not yet related to a clear structural variation~\cite{RevRoyall},
its fingerprint remains essentially of a dynamic type, hidden in the way the system moves.
\textgx{Indeed, in liquids close to the glass transition, DHs emerge as transient clusters of particles with a mobility larger or smaller than the average~\cite{WeeksScience}.
The size and the lifetime of these dynamical clusters increase on approaching the transition, playing a role similar to density fluctuations close to an ordinary critical point~\cite{Halperin, BBreview, Whitelam}.}
This motivated the glass community to develop a robust framework for characterizing DHs.
In glass forming liquids, the structural relaxation process as a function of time, $\Delta t$,
can be monitored through a dynamic order parameter
probing the local motion on the length scale of the particle size. 
Different experimentally measured probes, 
such as the dynamic scattering functions or the persistence, are good choice as order parameter and provide similar information~\cite{BerthierPhys}. 
The fluctuations of the dynamic order parameter define a dynamic susceptibility, $\chi_4(\Delta t)$, that allows for quantifying the degree of DH.
Alternatively, $\chi_4(\Delta t)$ is also defined as the space integral of a correlation function, 
$G_4(r, \Delta t)$, measuring correlations of the displacements over $\Delta t$ 
between particles separated by a distance $r$.  
These equivalent definitions of $\chi_4$ reveals the two faces of DHs, 
that can  be viewed either as ensemble fluctuations of the dynamic order parameter, 
or as spatial correlations in the displacement field~\cite{DHbook, Ediger}.

While direct evidences of DHs have been first provided by numerical simulations~\cite{Glotzer, Glotzer2},
their existence have been directly confirmed by experiments on colloidal glasses and 
other colloidal systems, such as gels and foams~\cite{CipellettiPRL04,CipellettiEPL, Gao, Trappe}
and, recently, even in epithelial cell tissues~\cite{Angelini, Fredberg, Cerbino_cells}.

In fact, DH characterization still remains a complex experimental task, 
typically handled by a limited number of specialized academic groups, 
since it requires to resolve the dynamics in space and time, and estimate deviations from the
average behavior. 
As far as individual particle can be resolved, optical or confocal microscopy, combined with particle tracking 
allows for properly monitoring the macroscopic dynamics, measuring the dynamic order parameter
as well as other complex dynamic correlation functions, such as the bond-orientational correlation function~\cite{Vivek}.
This approach has been also exploited to obtain indirect and visual insights on DHs in colloidal systems~\cite{WeeksScience}.
\textgx{Particle tracking based quantification of DHs has been, instead, mainly performed in granular systems of large, non-thermal particles~\cite{Dauchot, Keys, Abate_Durian}. 
In analogy with numerical simulations~\cite{Lacevic}, a dynamic susceptibility, $\chi_4(l,\Delta t)$, is measured from the fluctuations of the fraction of particles
which moved more than an arbitrary chosen cutoff distance, $l$, over the time, $\Delta t$~\cite{Dauchot, Keys}. 
Alternatively, the cutoff distance can be determined uniquely by the topology,
for example by considering the fraction of sample which remains inside the
same Voronoi cell across $\Delta t$~\cite{Abate_Durian}.
However, these measurements are complicated and not always possible, especially in crowded colloids,}
since the length and the overall number of trajectories are often limited, even when particles are clearly resolved.
Moreover, particle tracking relies on quite complex algorithms and suffers possible biases due to users' choice of tracking parameters. 
By contrast, Dynamic Light Scattering (DLS) based techniques are probably the most robust approach to measure $\chi_4$~\cite{CipellettiNatPhys, Maggi},
but do not provide any direct visualization of DHs. 
Simultaneous visualization and quantitative measurement of  DHs have been obtained using more 
sophisticated techniques, such as the Photon Correlation Imaging (PCI), that combines features of both dynamic light scattering and imaging~\cite{CipellettiPRL09}.
Results on flowing systems suggest that $\chi_4$ can be measured using simpler methods, based
on autocorrelation of image intensity~\cite{Durian_SM, Durian_PRE}, and call for further exploring this direction.
\textgx{Recently, elegant approaches to investigate soft matter dynamics analyzing image
difference (the same signal exploited in this work) have been developed~\cite{Giavazzi, Cerbino}.
However, the current differential methods do not allow DH characterization. 
For instance,} Differential Dynamic Microscopic (DDM), that provides information similar to DLS from video microscopy data,
is an easy and promising technique~\cite{DDM, Helgeson}, but currently limited to monitor the structural relaxation~\cite{Lu, Zaccarelli} and not DHs.
It appears clearly that an easy way to characterize complex fluids with dynamic heterogeneity is highly desirable,
also considering that soft glassy materials are common in technological applications and biological systems.\\
In this paper, we introduce a novel and straightforward experimental method to fully characterize DHs in colloidal suspensions and apply it 
to a popular model system of hard-sphere colloidal glass~\cite{Pusey, Brambilla, Makse, Sood_NatPhys, Sood_NatComm} imaged by optical microscopy.
Our method utilizes as primary signal the {\it differential frames} obtained by subtracting images taken at different time.
This is also the signal used by DDM, before performing a Fourier analysis and accessing 
the intermediate scattering function by appropriate fitting of the image structure function~\cite{DDM}.
Our Differential Variance Analysis (DVA), instead, does not require Fourier analysis or fitting ansatzs. 
Indeed, we simply focus on the real space variance of differential frames and its fluctuations to
obtain the dynamic order parameter  and the dynamic susceptibility, respectively.
We validate the result of DVA by performing established single particle tracking analysis 
and demonstrating that the dynamic order parameter obtained from DVA closely matches the 
commonly measured Intermediate Self Scattering Function (ISSF) at a wave-length of the order of the particle size.
In addition, the  differential frames provides a very direct visualization of DHs:
the framework we introduce to this aim allows for visualizing DHs not only as spatial correlations, but also 
as ensemble fluctuations. 
\textgx{The key of this visualization is to consider differential frames close to the lag-time $\Delta t^{*}$,
which is determined by the dynamics and  can be easily measured from DVA as the time corresponding to the maximum of $\chi_4$.}

\section*{Results}
\subsection*{Differential Variance Analysis of glassy dynamics}
\label{res1}

\begin{figure}[ht]
\centering
\includegraphics[width=\linewidth]{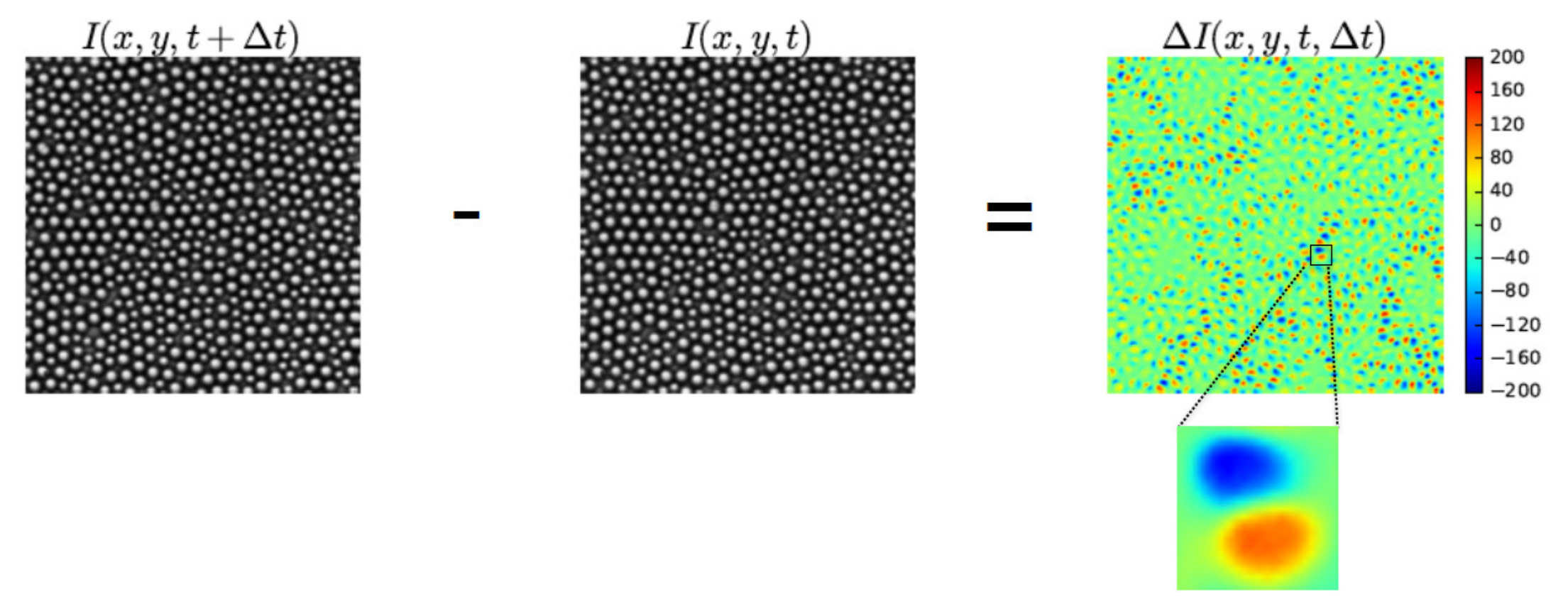}
\caption{\label{fig:line}
{\bf Differential frame.} 
From left to right, two snapshots of a portion of a sample separated by a lag-time $\Delta t=10 s$, and the resulting  differential frame. 
Particles that move significanlty during the interval $\Delta t$ give rise in the differential frames to coupled spots of high and low intensity, 
which look like dipoles, as highlighted 
by the zoom. The volume fraction of this sample is $\Phi=0.71$.}
\end{figure}

Data are obtained from a previous experiment~\cite{SM15_exp} that investigated a quasi two-dimensional mixture of micron-sized hard-sphere-like beads in water
(see Methods).
In this popular model system of colloidal glass, the dynamics slows down on increasing the colloidal volume fraction.
From optical video microscopy of these samples, we consider two frames at time $t$ and $t+\Delta t$, and the \textit{differential frame} generated
by the differences between their pixel intensities, $\Delta I(x,y,t,\Delta t)=I(x,y,t+\Delta t)-I(x,y,t)$, 
as illustrated in Figure~\ref{fig:line} for  two frames of a sample at $\Phi=0.71$ and separated by a lag-time $\Delta t=10 s$, 
somewhat smaller than  the structural relaxation time.
On this timescale, some particles move over a distance comparable to their size, while
other particles stay localized close to their initial position~\cite{WeeksScience}. Such a scenario clearly emerges from the differential frame.  
Indeed, a color scale for the differential intensity signal highlights the presence of patterns formed by two adjacent spots of 
negative and positive $\Delta I$, that appear blue and red, respectively. These spots arise as a consequence of detectable single particle movements:
a blue spot corresponds to groups of pixels which are occupied by a particle at time $t$ but not at time $t+\Delta t$, and vice-versa for a red spot. 
Thus, each pair of blue and red spot can be viewed as a dipole or as an arrow representing the particle displacement.
By contrast, the green background corresponds to regions occupied by particles that at time $t+\Delta t$ are still localized in their original position (at time $t$),
with thermal rattling around this position resulting in small  deviations from $\Delta I = 0$.
\begin{figure}[ht]
\centering
\includegraphics[width=12cm]{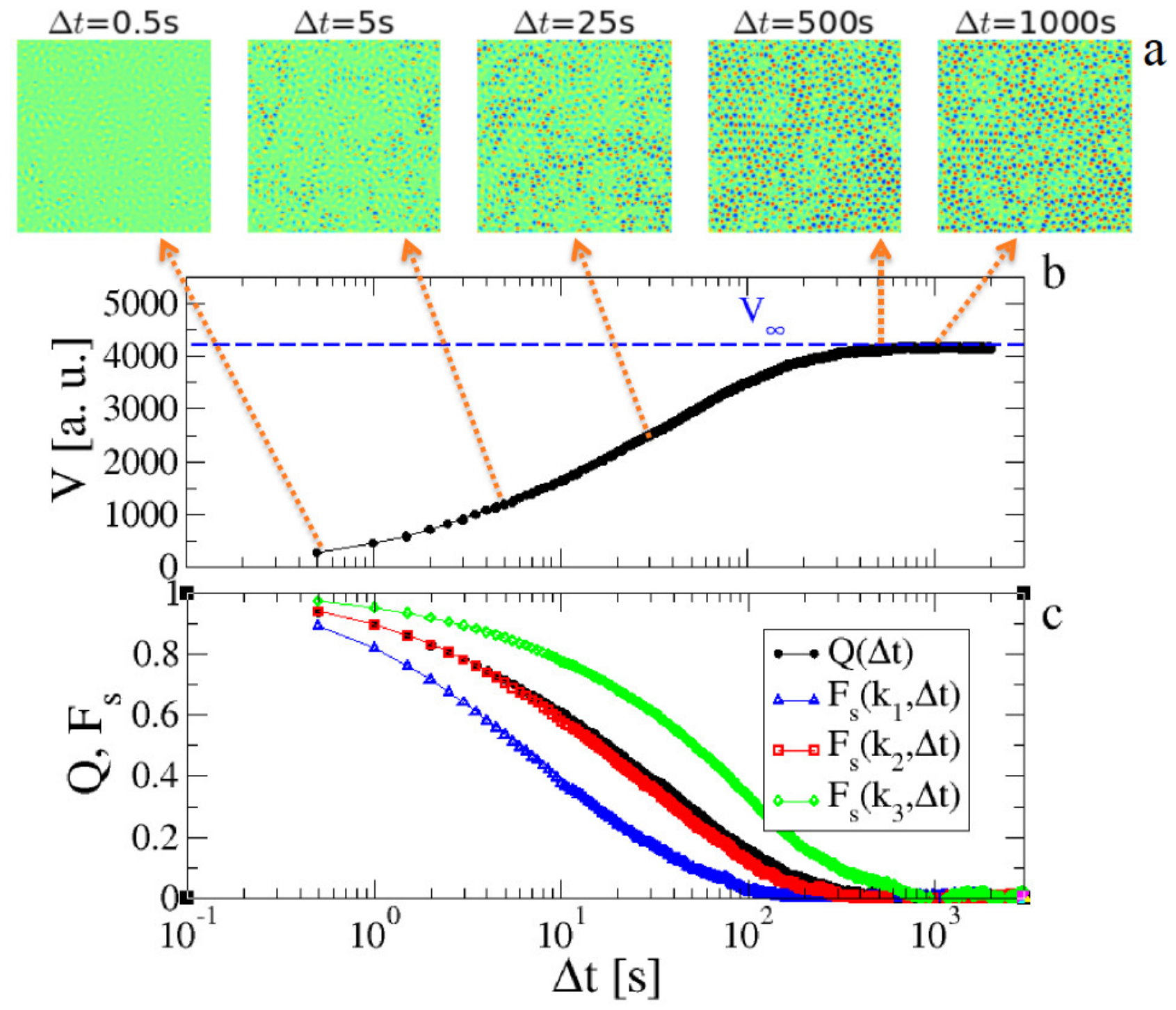}
\caption{\label{fig:overlap}
{\bf From differential frames to the dynamic order parameter.}
 a) Sequence of differential frames at different lag-times, $\Delta t$, as indicated.
 b) Average intensity variance of differential frames, $V$, as a function of the lag-time $\Delta t$. 
$V(\Delta t)$ increases up to reach a plateau $V_{\infty}$. 
b) The DVA dynamic order parameter, $Q(\Delta t)$, is compared to the ISSF,
$F_s(k,t)$, computed from single particle trajectories, for three wavectors $k_1=2.3 \mu m^{-1}$, $k_2=1.7 \mu m^{-1}$ and $k_3=1.15 \mu m^{-1}$, as indicated.
The volume fraction of the sample under consideration is $\Phi=0.71$. 
}
\end{figure}
To qualitatively illustrate the system temporal evolution, Figure~\ref{fig:overlap}a shows a sequence of these differential frames at increasing lag-times.
Initially, as $\Delta t$ increases, more and more particles move, leaving dipoles in the differential frames.
At lag-times much larger than the relaxation time, instead, all particles have moved far away from from their original position and the number of dipoles seems to saturate.
Quantitatively, this temporal evolution can be captured by the variance of $\Delta I$ over pixels ($x,y$):
\begin{equation}
\label{eq:V}
\hat V(t, \Delta t)=\frac{1}{L^2} \int_{L^2} dx dy \Delta I^2(x,y,t,\Delta t),
\end{equation}
where $L$ is the size of the image.
Figure~\ref{fig:overlap}b shows $V(\Delta t)=\langle \hat V(t, \Delta t) \rangle_t$, obtained by averaging $\hat V(t, \Delta t)$ over an ensemble of differential frames with different initial times $t$.
A comparison with Figure~\ref{fig:overlap}a, which refers to the same volume fraction, $\Phi=0.71$,
clarifies that in the time window in which the number of dipoles increases,  $V (\Delta t)$ also increases, whereas  it approaches a plateau, $V_{\infty}$, at long time,
when the number of dipoles saturates. This suggests that the behaviour of the variance is closely related to the relaxation process.
Indeed, we are going to show that the average and the fluctuations of $\hat V (\Delta t)$  
can be used to quantitatively describe the structural relaxation and the emergence of dynamic heterogeneities, respectively.
To this aim, we introduce the function
\begin{equation}
\label{eq:overlap}
\hat Q(t, \Delta t)=1-\frac{\hat V(t, \Delta t)}{V_{\infty}},
\end{equation}
and speculate that its average, $Q(\Delta t)=\langle \hat Q(t, \Delta t)\rangle_t$, properly describes the structural relaxation.
To demonstrate this point, we measure  the commonly used ISSF,
 $F_s(k,\Delta t)= \langle \Psi_t(k,\Delta t) \rangle_t$, 
where $\Psi_t(k,\Delta t)=\frac{1}{N}\sum_{i=1} ^N e^{-ik[r_i(t+\Delta t)-r_i(t)]}$, $N$ is the number of particles under consideration, 
and the trajectories $r_i(t)$ are obtained trough single particle tracking~\cite{tracking1, tracking2}.
Figure~\ref{fig:overlap}c shows $Q(\Delta t)$,
obtained from the variance in panel a, and $F_s(k,\Delta t)$ for three values of the wave-vector, $k=2\pi / \lambda$,
selected in a range relevant to describe the structural relaxation:
$k_1$, $k_2$ and $k_3$,  corresponding to wave-lengths $\lambda_1=d$, $\lambda_2=1.35 d$, and $\lambda_3=2 d$, respectively, 
with $d\simeq2.7 \mu m$ being the average particle diameter.
Strikingly, $Q(\Delta t)$ lies between $F_s(k_1,\Delta t)$ and $F_s(k_3,\Delta t)$ and
nearly overlaps to $F_s(k_2,\Delta t)$ at any time.~\cite{nota1}

\begin{figure}[ht]
\centering
\includegraphics[width=\linewidth]{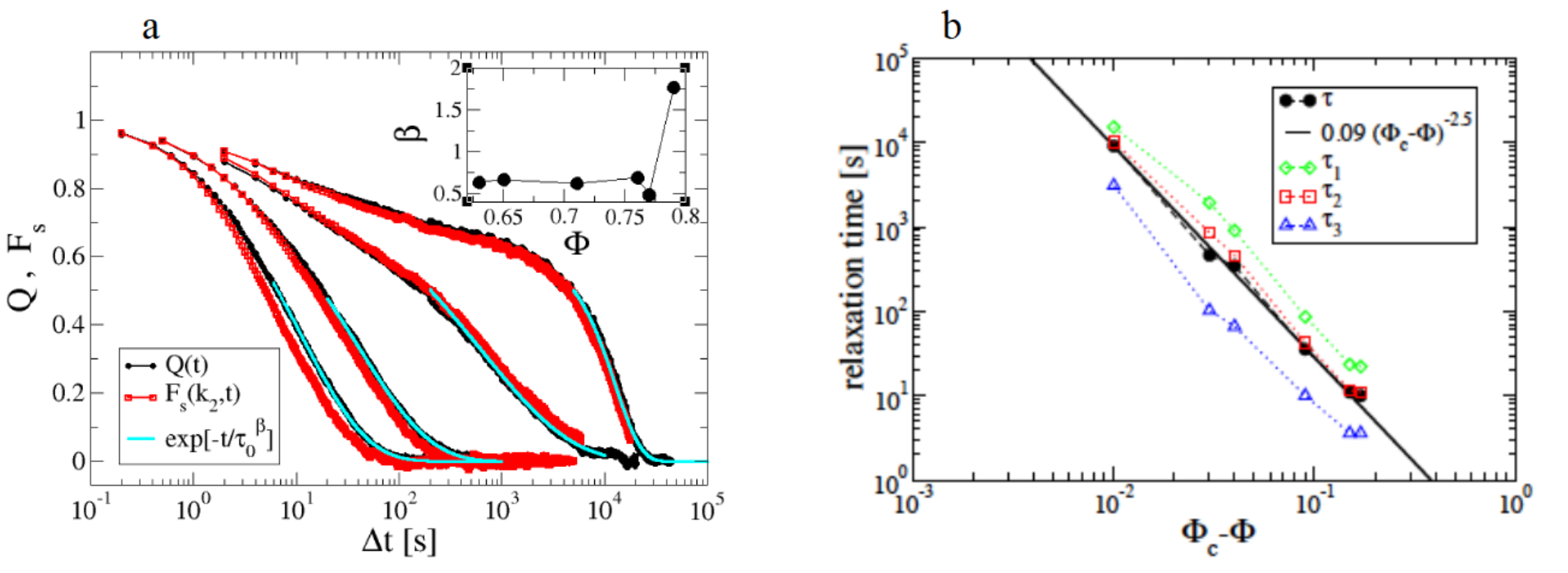}
\caption{\label{fig:structural_relaxation} 
{\bf Structural relaxation.}
a) DVA dynamic order parameter, $Q(\Delta t)$, and ISSF, $F_s(k_2,\Delta t)$, with $k_2=1.7 \mu m^{-1}$ at volume fractions $\Phi=0.65, 0.71, 0.77, 0.79$, from left to right.
$Q(\Delta t)$ is nearly overlapped to $F_s(k_2,\Delta t)$ over the whole range of time and investigated volume fractions. 
The solid lines are fits, $Ae^{-{t/\tau_0}^{\beta}}$,  to the late decay of $Q(\Delta t)$. 
The resulting exponent $\beta$ as a function of $\Phi$ is reported in the Inset.
b) Relaxation time measured from the $Q$ decay, $\tau$, and from the ISSF decays, $\tau_1$, $\tau_2$, $\tau_3$  
(corresponding to wave-vectors $k_1$, $k_2$ and $k_3$, respectively), as a function of $\Phi_c-\Phi$, with $\Phi_c=0.8$.
$\tau$ is well fitted by a power-law $(\Phi_c-\Phi)^{-2.5}$ (solid line). 
The ISSF relaxation times follows a similar behaviour, with $\tau_2$ being nearly overlapped to $\tau$.
}
\end{figure}

In addition, Figure~\ref{fig:structural_relaxation}a shows that the close similarity between $Q(\Delta t)$ and $F_s(k_2,\Delta t)$ 
is manifested over the whole range of investigated volume fractions. 
Overall, these results clarify that $Q(\Delta t)$ is an effective dynamic order parameter of the system structural relaxation, 
probing a length scale of the order of one particle diameter. 
\textgx{This length scale is not arbitrary, but self-determined by the local structure, in analogy with Ref.~\cite{Abate_Durian}.}

We note as an aside that the late decay of $Q(\Delta t)$ is well fitted by the functional form $Ae^{-{t/\tau_0}^{\beta}}$ (solid lines),
as usually found in glassy materials. The estimated value of the exponent poorly fluctuates around $\beta\simeq0.55$ 
at all the volume fractions investigated, except for the largest one, where it jumps to $\beta\simeq 1.75$ (see Inset of Figure~\ref{fig:structural_relaxation} a). 
A similar crossover from stretched ($\beta<1$) to compressed exponential ($\beta>1$) has been previously reported in nearly hard sphere colloidal glasses~\cite{CipellettiNatPhys}
as well as in other glassy systems, and its origin is currently debated~\cite{Delgado, Ferrero}.  

The decay of $Q(\Delta t)$ allows for estimating the structural relaxation time, $\tau$,  from the relation $Q(\tau)=1/e$.
Figure~\ref{fig:structural_relaxation}b shows that the dependence of $\tau$ on the volume fraction is compatible with a power-law, $(\Phi-\Phi_c)^{-\gamma}$,
as predicted  by Mode Coupling Theory (MCT). In particular, we find  $\Phi_c\simeq0.80\pm0.01$ and $\gamma=2.5\pm0.1$.
The figure also shows the relaxation times of the ISSFs, clarifying that the results of Figs.~\ref{fig:overlap}b and~\ref{fig:structural_relaxation}a  
are fully reflected in the relaxation times. 
Indeed, $\tau$ is nearly overlapped to the relaxation time of $F_s(k_2,\Delta t)$ and in between those
of  $F_s(k_1,\Delta t)$ and $F_s(k_3,\Delta t)$, all the relaxation times being compatible with the same power-law.

The emergence of dynamic heterogeneity can be now characterized defining the dynamic susceptibility from the fluctuations of the DVA
dynamic order parameter, $Q$:
\begin{equation}
\label{eq:chi4_3}
\chi_4(\Delta t)=N\left [ \langle \hat Q^2(t, \Delta t) \rangle_t - \langle \hat Q(t, \Delta t) \rangle_t^2 \right ]
\end{equation}
which is directly related to fluctuations of the variance using eq.~\ref{eq:V}, 
$\chi_4(\Delta t)=\frac{N}{V_{\infty}^2} \left [\langle \hat V^2(t, \Delta t) \rangle_t - \langle \hat V(t, \Delta t)\rangle_t^2 \right ]$.

\begin{figure}[ht]
\centering
\includegraphics[width=\linewidth]{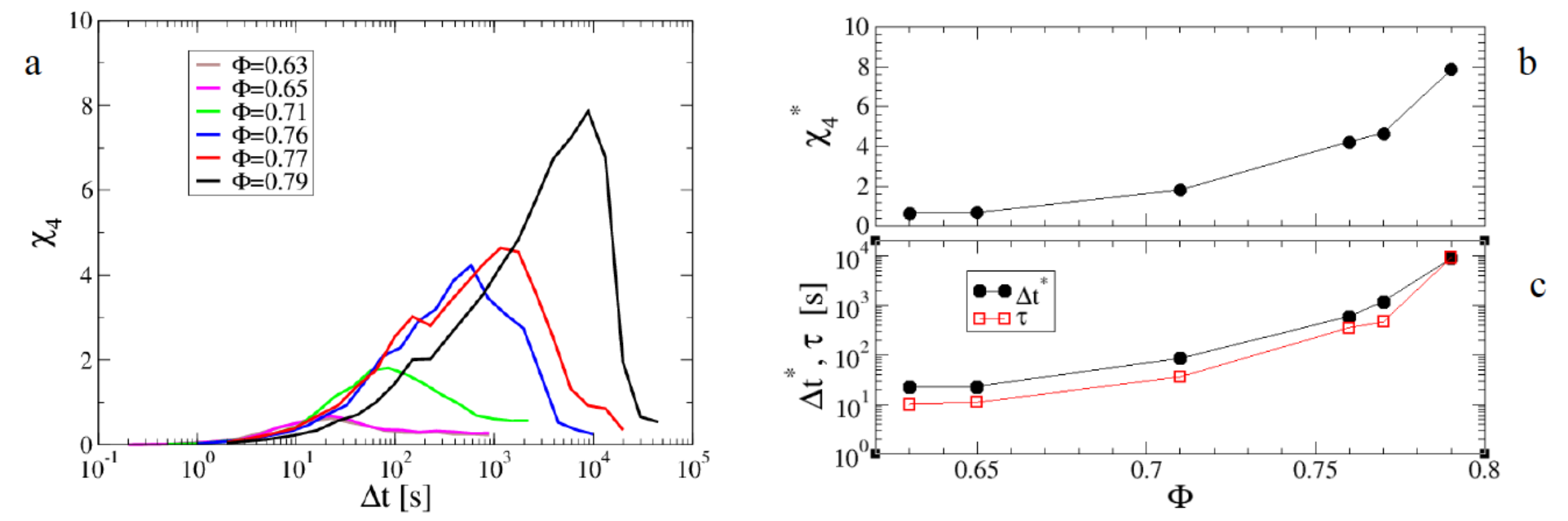}
\caption{\label{fig:chi4}
{\bf Quantification of Dynamic Heterogeneities.}
a) Dynamic susceptibility, $\chi_4$, as function of the lag-time $\Delta t$ for different volume fractions, as indicated.
$\chi_4(\Delta t)$ shows a maximum $\chi_4^*$ at a time $\Delta t^*$.
b) $\chi_4^*$  as a function of the volume fraction, $\Phi$.  
c) $\Delta t^*$ as a function of $\Phi$ is compared to the relaxation time, $\tau$.  
}
\end{figure}

Figure~\ref{fig:chi4}a shows that $\chi_4(t)$ has the typical behaviour  
reported for the dynamic susceptibility in glass-formers~\cite{DHbook}, with a maximum $\chi_4^*$ at a time $\Delta t^*$,
both clearly increasing on increasing the volume fraction.
$\chi_4^*$ and $\Delta t^*$ roughly estimate the typical size and life-time of clusters of particles dynamically correlated.
Accordingly, the mentioned similarities with ordinary critical phenomena emerge since these dynamical clusters 
become increasingly spatially extended and long-living on approaching the glass transition~\cite{DHbook}.
Figure~\ref{fig:chi4}b and c shows that for this system $\chi_4^*$ increases about a decade over the investigated range of volume fractions,
while $\Delta t^*$ spans almost three orders of magnitude, roughly mimicking the behaviour of the relaxation time, $\tau$.  
Let us stress that  DVA allows for a simple and efficient measure of the dynamic susceptibility,
since it does exploit the whole statistics provided by the raw video microscopy data and is directly applicable, 
without preprocessing the images or resolving individual particle positions.
Using single particle tracking, instead, we can provide a reliable measurement of the ISSF, but  not of the associated susceptibility,
$\chi_4=N[ \langle \Psi_t(k,\Delta t)^2 \rangle_t -  \langle \Psi_t(k,\Delta t) \rangle_t^2]$.
Indeed, the ISSF can be computed averaging over all the tracked particles, no matter the initial and the final time of each trajectory,
whereas computing the associated fluctuations, and in particular the square average, $\langle \Psi_t(k,\Delta t)^2 \rangle_t $,
requires the trajectories to be temporally overlapped during time-windows much larger than the relaxation time.
As this time-window increases, more and more trajectories are rejected 
 since particles can diffuse away from the field of view or due to incidental failure of the tracking algorithm.
This strongly limits the number of available trajectories,
especially at high density, where the relaxation time is large,  
and large statistics should be required to properly estimate $\chi_4$.
For example, at the largest volume fraction, we are able to record only a few tens trajectories that are both temporally overlapped and longer than $\tau$.


\subsection*{Direct visualization of Dynamic Heterogeneities} 
\label {res2}
\begin{figure}[ht!]
\centering
\includegraphics[width=15cm]{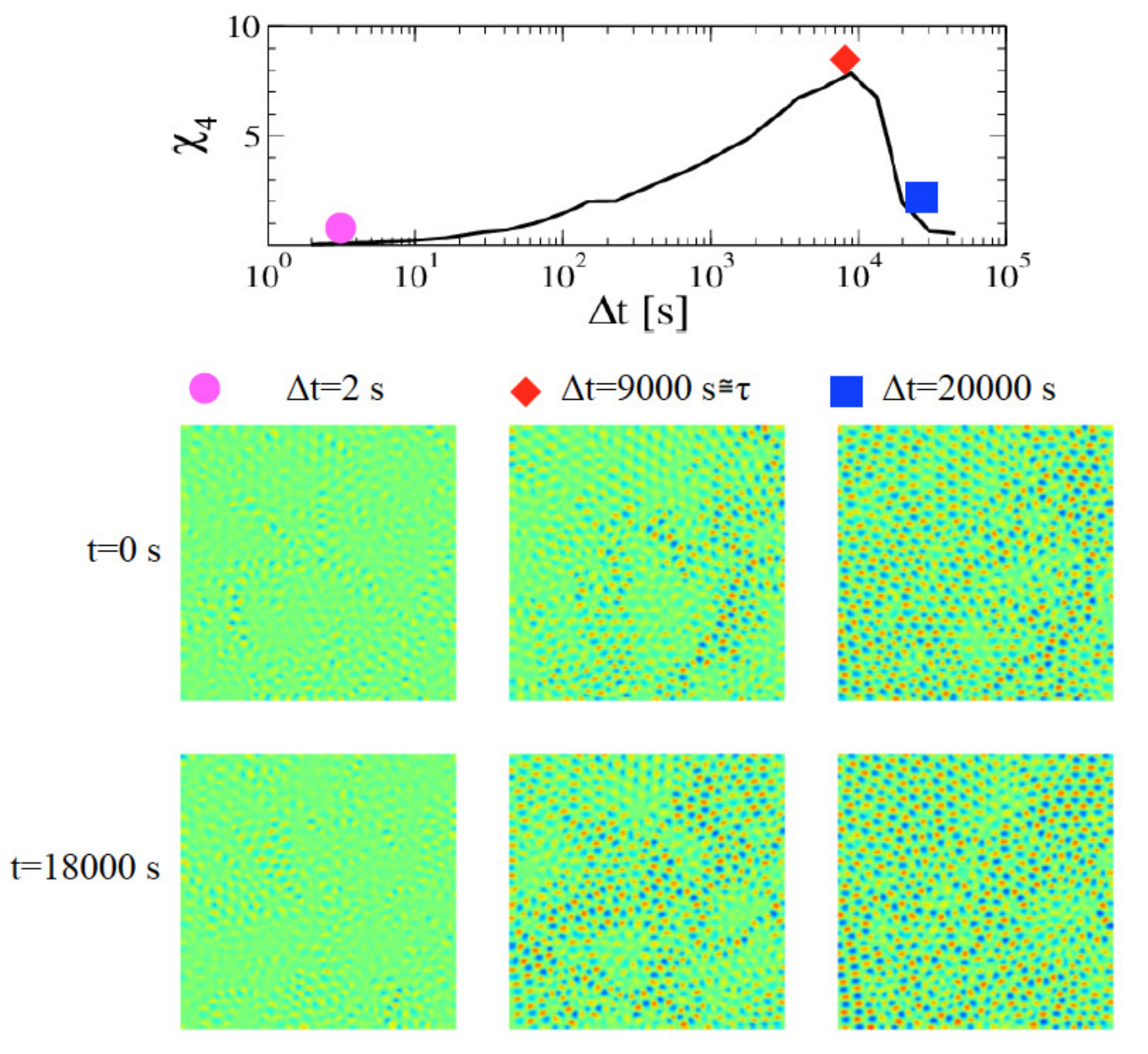}
\caption{\label{fig:matrix} 
Visualization of Dynamic Heterogeneities.
Matrix of differential frames with different initial times $t$ and lag-times $\Delta t$, 
for a portion of a sample at the largest investigated volume fraction, $\Phi=0.79$.
For each line, $t$ is fixed  and $\Delta t$ increases moving from left to right, as indicated.
The first and the third frame of each line report values of $\Delta t$ much smaller and much larger than the relaxation time, $\tau$,
 while $\Delta t\simeq \tau\simeq \Delta t^*$ for the second one.
The two lines refers to different initial times $t$, separated by a delay of the order of $2\tau$.
At $\Delta t\simeq\tau$, dynamic heterogeneities are manifested, either in each single frame as large spatial correlations among the dipoles corresponding to the moved particles,
or as fluctuations of the number of dipoles between the two lines. 
In order to have a comparison with quantitative measure of DHs, the upper panel reports $\chi_4(\Delta t)$ for the same volume fraction. 
The symbols highlights the values of $\chi_4$ at the lag-times reported in the matrix.
} 
\end{figure}
Now we turn back to direct observation of differential frames,
in order to show how this approach naturally leads to a novel and effective visualization of DHs.
To this aim, Figure~\ref{fig:matrix} (lower panel) shows a matrix of differential frames at the largest investigated volume fraction, $\Phi=0.79$.
Moving along a line, the initial time,  $t$, is fixed, while the lag-time, $\Delta t$, is increasing and 
the system is progressively relaxing with respect to the initial configuration.
\textgx{Note that the value of $\Delta t$ corresponding to the central frame is chosen not arbitrarily,
but close to the lag-time, $\Delta t^*$, of the maximum $\chi_4$, which for this systems is also of the order of the relaxation time, $\tau$.}
The two lines of the matrix correspond to different initial times $t$. 
Since these $t's$ are separated by a time larger than the relaxation time, $\tau$,
the corresponding configurations are uncorrelated and, therefore, akin to different replicas of the same system, as commonly  generated in numerical simulations. 
For a comparison with quantitative results,  the upper panel reports $\chi_4(\Delta t)$ 
at the same volume fraction, highlighting its value  at the lag-times considered in the matrix below.
The Figure shows that in each frame at $\Delta t \simeq \Delta t^* \simeq \tau$ (second column), 
dipoles corresponding to moved particles exhibit large spatial correlations, and coexist
with extended frozen swarms ($\Delta I\simeq0$), where the system has not yet relaxed.
The emerging picture resembles the dynamic phase coexistence scenario, which ascribes 
the glassy dynamics to the temporary coexistence of a mobile/liquid and an immobile/solid phase~\cite{SciRep}. 
Conversely, the differential frames look quite homogeneous at short and long time (first and third columns).
Overall, this is reflecting the quantitative informations provided by  $\chi_4(\Delta t)$ which is maximum around $\Delta t \simeq \tau$,
and small at shorter and longer time (see the upper panel).
Furthermore, comparing the two lines of the matrix allows for an alternative visualization of DHs,
which become manifested in the fluctuations of dipole patterns between differential frames at the same $\Delta t$, but different $t$. 
Once again, these fluctuations are marked for differential frames at $\Delta t \simeq \tau$: it appears clearly that 
a much smaller fraction of the system  has relaxed in the upper frame compared to the bottom one,
despite that the considered lag-time is the same in the two frames. 
By contrast, fluctuations are negligible at short and long time.
In general, these signatures of DHs become less evident at smaller $\Phi$, where the maximum of $\chi_4$ is also smaller.

Observation of differential frames  at $\Delta t\simeq \tau$ suggests other interesting features of DHs:
for example, dynamical clusters of close dipoles resemble a correlated percolation patterns~\cite{Harrowell, Mallamace, PRL2011, Fractals, Chaudhuri}.
In addition, the shape of these clusters looks more compact than that observed at lower volume fraction 
(see for example the third frame in Figure~\ref{fig:overlap}a, where $\Delta t$ is also of the order of the relaxation time at that volume fraction),
consistently with a string to compact crossover of cooperative rearrangements on approaching the glass transition~\cite{Woolines, Sood}. 
 
 \section*{Discussion}
The results of the previous Section demonstrate the ability to quantitatively monitor the structural relaxation process and the emergence of DHs, 
starting from the variance of the differential frames.
In particular, {\it( i)} we introduced a dynamic order parameter, $Q(\Delta t)$, that properly describes the relaxation process,
as demonstrated by a comparison with the commonly used ISSF, 
{\it (ii)}  we measured the structural relaxation time, $\tau$, from the decay of $Q(\Delta t)$, 
and finally {\it (iii)} the dynamic susceptibility $\chi_4(\Delta t)$ from its fluctuations.\\
Moreover, this method leads to directly visualizing the relaxation process and the emergence of DHs.
In particular, Figure~\ref{fig:matrix} summarizes  what is the key of our approach:
as the time $\Delta t$ passes, an increasing number of well defined dipoles appears, 
which controls the behaviour of the variance, and, therefore, the dynamic order parameter and its fluctuations. 
Indeed, these dipoles are the signature of particle rearrangements that leads to the structural relaxation and are of order of one particle diameter,
since this is the length scale probed by $Q(t)$.
Considering the strongly intermittent nature of particle motion in glasses~\cite{WeeksScience, SM_review},
it is likely that variations in the differential frames with $\Delta t$ are mainly due 
to the increase of the number of dipoles, rather than to change of their intensity.
Accordingly, we have seen that, focusing on dipole patterns, DHs naturally emerge in the differential frames.
Incidentally, we mentioned that DHs have a double-sided nature:
they are manifested both as spatial correlations in the particle displacements and ensemble fluctuations of the dynamic order parameter.
Accordingly, the dynamic susceptibility can be equivalently defined from {\it (I)} the space integral of $G_4(r, \Delta t)$
or from {\it  (II)} the ensemble fluctuations of the dynamic order parameter, like in eq.~\ref{eq:chi4_3}.
However, to measure $\chi_4$ in practice, {\it (II)} is largely preferred to {\it (I)}.
Experimental techniques, such as DLS, do not give information on the particle positions,
but only on the dynamic order parameter and its fluctuations. 
Even in simulations, where the particle positions can be fully resolved,
{\it (I)} is poorly used in practice, due to the difficulty in obtaining reliable measurements of $G_4(r,\Delta t)$~\cite{DHbook}.
Nevertheless, correlations in the particle displacements provide important qualitative insights,  being at the base of DH direct visualization proposed until now,
for example highlighting the position of the "fast particles", i.e. the particles which have moved more than a given threshold over a time interval of the order of the relaxation time~\cite{WeeksScience}.
We have seen that our approach leads to a similar goal easily, since such an information is already contained in the raw differential frames at $\Delta t\simeq \Delta t^* \simeq \tau$:
particles that have moved significantly give rise to a dipoles and DHs becomes apparent as spatial correlations among these dipoles.
In addition, DHs are also manifested as large fluctuations of the number of dipoles between differential frames at lag-time  $\Delta t\simeq \Delta t^* \simeq \tau$,
and different initial times, $t$.
This latter is a way to directly visualize the ensemble fluctuations of the dynamic order parameter {\it  (II)}, which are actually used to compute $\chi_4$. 

While previous efforts used an arbitrary fixed lag-time~\cite{Scheffold}, 
we remark the importance of choosing a well defined timescale, namely $\Delta t^*$, to effectively visualize DHs.
This timescale is determined by the dynamics and, therefore, changes on varying the system control parameter ($\Phi$ in our experiment).

\section*{Conclusions}
In this paper, we introduced DVA as a novel and simple method to characterize the dynamics of hard-sphere colloidal glasses
of micron-sized particles. We expect DVA to be applicable to a large range of experimental systems formed by different colloidal particles,
such as soft particles or attractive particles, which likely form gels, as well as red blood or epithelial cells, 
and, with three dimensional systems imaged by confocal microscopy.
These experimental systems are very popular and a large amount of imaging data
have been collected during the last years and mainly analysed by single particle tracking.
Previously recorded videos can be easily reprocessed utilizing this approach to obtain information, complementary to that provided
by particle tracking and an effective direct visualization of the heterogeneous dynamics.
Moreover, preliminary results suggest that DVA could be also suited to
systems formed by much smaller (in the nanometer range) primary particles.

Understanding whether the heterogeneous dynamics in glasses had a structural origin is still one of the most relevant open issue
in condensed matter physics~\cite{RevRoyall, Tanaka}. Indeed, the presence of structural heterogeneities implies that of DHs but the opposite is not true~\cite{BerthierPhys}.
Accordingly, DHs are predicted by  several theoretical scenarios both
postulating a structural~\cite{RFOT} or a purely dynamic origin~\cite{Facilitation} for the glass transition.
By contrast, the heterogeneous dynamics of other materials, like gels or fiber networks, is known to have a structural origin.
Yet, in practice,  structural characterization of these materials requires quite complex experimental efforts,
and inferring structural informations from the dynamics can be often an easier alternative.
Since these systems are widespread in industry and biology, our method
could be very convenient to control their degree of heterogeneity,
focussing on the dynamics.

Finally, we suggest that the DVA strategy could be interestingly extended to a wide variety of data sets, not only to video microscopy and imaging data.   


\section*{Methods}

\label{Sec:methods}

Data were obtained from previous experiments~\cite{SM15_exp}, which investigated the intermittent single particle motion using particle tracking. 
The investigated systems was a quasi-two dimensional hard-sphere-like colloidal systems at different volume fractions, $\Phi$.
Precisely, the samples consisted in a 50:50 binary mixture of silica beads dispersed in a water surfactant solution (Triton X-100, 0.2\% v/v),
to avoid particle sticking through van der Waals forces. 
Large and small beads had diameters 3.16 and 2.31 $\mu m$, respectively, resulting in an $1.4$ ratio known to prevent crystallization.
The system were imaged using a standard microscope equipped with a 40x objective (OlympusUPLAPO 40XS) and 
the images were recorded using a fast digital camera (Prosilica GE680). At the highest volume fraction, 
roughly a thousand particles in the field of view of the microscope were imaged.
We focused on a volume fraction range, where the samples can be equilibrated on the experimental timescale and monitored
the dynamics after thermal equilibrium is attained. 
Videos recorded at each volume fraction were several times larger than the relaxation time, $\tau$.
In particular, the video durations and frames per seconds (fps) ranged in $[10^3 s, 10^5 s]$ and $[0.5 s^{-1}, 5 s^-{1}]$, respectively, 
depending on the volume fraction, i.e. larger duration and smaller fps at larger volume fraction.
Data analysis was performed using Python and different SciPy libraries~\cite{Python1}.
Interactive data exploration and visualization was performed using IPython and Jupyter notebooks~\cite{Python2}.
DVA code is freely available on the corresponding author web-page, \url{http://rpastore.altervista.org/DVA/}.

\section*{Acknowledgements}
We are  indebted to D. Bolchini, A. Fierro, F. Giavazzi, V. Guida, S. Guido, G. Raos, M. Ribera d\'~Alcal\'a, E. Santamato, V. Trappe,
and deeply indebted to A. Coniglio and R. Cerbino for many useful discussions and critical reading of the manuscript.

We thank The Procter and Gamble Co. and UC Simulation Center for supporting 
and funding R.P. as visiting researcher in Cincinnati during the last months.
This research is also supported by the SPIN SEED 2014 project {\it Charge separation and
charge transport in hybrid solar cells} and the CNR--NTU joint laboratory {\it Amorphous materials for energy harvesting applications}.

\section*{Author Contributions}
R.P. and M.C. conceived the project and designed DVA. 
G.P. performed the experiments and the single-particle tracking. 
R.P. and M.C. performed data analysis and interpretation. 
R.P. wrote the paper with the contribution of the other authors.

\section*{Additional information}
The authors declare no competing financial interests.


\begin{thebibliography}{30}
\bibitem{nota1} While further connections with wave-vector dependent dynamic correlators will be  the object of an upcoming work,
it is worth noticing that the similarity between $F_s(k_2,t)$ and $Q(t)$ 
can be rationalized focusing on the Fourier transform of the differential frame intensity 
which in our system always shows a maximum around $k_2$.
\end{thebibliography}
\end{document}